\documentclass[sigconf]{acmart}

\renewcommand\footnotetextcopyrightpermission[1]{} 
\AtBeginDocument{%
  }

\copyrightyear{2025}
\acmYear{2025}
\setcopyright{cc}
\setcctype{by}
\acmConference[CIKM '25]{Proceedings of the 34th ACM International Conference on Information and Knowledge Management}{November 10--14, 2025}{Seoul, Republic of Korea}
\acmBooktitle{Proceedings of the 34th ACM International Conference on Information and Knowledge Management (CIKM '25), November 10--14, 2025, Seoul, Republic of Korea}\acmDOI{10.1145/3746252.3761651}
\acmISBN{979-8-4007-2040-6/2025/11}

\usepackage{appendix}
\usepackage{tabularx}
\usepackage{cleveref}
\usepackage{siunitx}
\usepackage{pgfplots}
\pgfplotsset{compat=1.18}
\usepackage{tikz}
\usepackage{graphicx}
\usepackage{multirow}
\usepackage{supertabular}
\usepackage{glossaries}

\newacronym{foltr}{FOLTR}{Federated Online Learning to Rank}
\newacronym{oltr}{OLTR}{Online Learning to Rank}
\newacronym{ltr}{LTR}{Learning to Rank}

\renewcommand\footnotetextcopyrightpermission[1]{}
\usepackage{fancyhdr}
\fancypagestyle{firstpage}{
  \fancyhf{}
  
  \fancyfoot[C]{\footnotesize This manuscript was accepted at CIKM 2025.
  The final version will be published in the ACM Digital Library.}
}

\begin{document}

\title[A Large-Scale Web Search Dataset for Federated Online Learning to Rank]{A Large-Scale Web Search Dataset for\\Federated Online Learning to Rank}


\author{Marcel Gregoriadis}
\email{m.gregoriadis@tudelft.nl}
\orcid{0000-0001-5094-0111}
\affiliation{%
  \institution{Delft University of Technology}
  \city{Delft}
  \country{The Netherlands}
}

\author{Jingwei Kang}
\email{j.kang@uva.nl}
\orcid{0009-0003-9283-4060}
\affiliation{%
  \institution{University of Amsterdam}
  \city{Amsterdam}
  \country{The Netherlands}
}

\author{Johan Pouwelse}
\email{j.a.pouwelse@tudelft.nl}
\orcid{0000-0002-9882-1506}
\affiliation{%
  \institution{Delft University of Technology}
  \city{Delft}
  \country{The Netherlands}
}

\renewcommand{\shortauthors}{Gregoriadis et al.}

\begin{abstract}
\glsresetall
  The centralized collection of search interaction logs for training ranking models raises significant privacy concerns.
  \gls{foltr} offers a privacy-preserving alternative by enabling collaborative model training without sharing raw user data.
  However, benchmarks in \gls{foltr} are largely based on random partitioning of classical learning-to-rank datasets, simulated user clicks, and the assumption of synchronous client participation.
  This oversimplifies real-world dynamics and undermines the realism of experimental results.
  We present AOL4FOLTR, a large-scale web search dataset with $\approx2.6$ million queries from \num{10000} users.
  Our dataset addresses key limitations of existing benchmarks by including user identifiers, real click data, and query timestamps, enabling realistic user partitioning, behavior modeling, and asynchronous federated learning scenarios.
\end{abstract}




\maketitle

\thispagestyle{firstpage}

\section{Introduction}
\glsresetall
\gls{oltr} is a widely used technique that aims to learn a ranker from users' interactions with search results.
The centralized data collection, however, exposes users to privacy risks, as query and interaction logs reveal sensitive information, such as demographic attributes or political views~\cite{barbaro2006face,sousa2021privacy,carpineto2016review}.
Federated learning approaches have been explored to address user privacy concerns~\cite{wang2021efficient,kharitonov2019federated,wang2021effective,wang2021federated}.
In \gls{foltr}, clients train a local ranking model on their personal interactions with search results, and collaboratively update a global model via privacy-preserving protocols.
While \gls{foltr} presents a promising approach for developing privacy-preserving ranking models, its evaluation is constrained by the absence of publicly available datasets.
In order to simulate client behavior,
existing work relies on random partitioning of classical offline learning-to-rank datasets~\cite{wang2021effective}, and the simulation of user interactions based on click models~\cite{kharitonov2019federated,wang2021effective,wang2021federated}.
This is inadequate as users have different document and click preferences~\cite{wang2022non,cecchetti2024learning}, giving rise to the non-IID problem in federated learning~\cite{zhu2021federated}.
Clients also vary in usage frequency, i.e., the data quantity they contribute to the global model.
As \citet{wang2022non} verified, this heterogeneity in client data poses a threat to \gls{foltr}, as models learn less effectively.
Moreover, existing work considers synchronous federated learning settings, which are inflexible and do not scale~\cite{yang2020federated}.
Realistically, individual client updates arrive with varying frequency and burst patterns.
This further impedes model convergence through issues related to staleness~\cite{xie2019asynchronous} and fairness~\cite{mohri2019agnostic}.
Despite the heterogeneous nature of real client data, and the asynchronicity of search interactions, \gls{foltr} is commonly simulated in synchronous settings and with IID data.
Accurate simulations demand a dataset with real user profiles and query timestamps.
Existing datasets typically aggregate data across large user populations to preserve individual privacy~\cite{craswell2020orcas,qin2013introducing}.

In this work, we present AOL4FOLTR~\cite{gregoriadis_2025_15678398}, the first real-world dataset for \gls{foltr}.
It contains more than 2.5 million search interactions from 10 thousand users, including raw queries and documents, user IDs, timestamps, clicked and non-clicked documents (i.e., result lists).
We base our dataset in the AOL query logs released in 2006~\cite{brenes2009stratified}.
We scraped the original website content at query time using the Internet Archive, recovering more than 420 thousand websites.
Furthermore, we used this collection of websites as a basis for reconstructing top-20 result sets for each query.
Finally, we encoded query-document pairs using 103 features, following conventions from popular learning-to-rank datasets.
We believe that our dataset positions itself as an important baseline for evaluating both synchronous and asynchronous \gls{foltr} scenarios.
Our dataset and code are made
publicly available, along with documentation to facilitate reproducibility\footnote{\url{https://github.com/mg98/aol4foltr}}.

\section{Background and Related Work}\label{sec:related}


Traditional (offline) \gls{ltr} datasets include
Microsoft's WEB10k/30k~\cite{qin2013introducing}, 
as well as datasets from Yahoo~\cite{chapelle2011yahoo} and Istella~\cite{dato2022istella22}. 
In these datasets, each query-document pair is annotated by humans with relevance scores from 0 (irrelevant) to 4 (highly relevant), which is used to optimize the ranking model.
%
In \gls{oltr}, the ranking model is continuously updated using real-time user clicks, which serve as implicit signals.
Unfortunately, there is no publicly available LTR dataset containing real user interaction data (i.e., click/no-click)~\cite{wang2021federated}. 
As a result, researchers often resort to LTR datasets with simulated user interactions generated by click models \cite{oosterhuis2018differentiable,grotov2016online,hofmann2013reusing}.
This approach is also commonly found in the evaluation of \gls{foltr}~\cite{kharitonov2019federated,wang2021effective,wang2021federated}.
In \gls{foltr}, clients train local ranking models on local data, and occasionally synchronize with a global model on a centralized server, where client updates are aggregated through methods like FedAvg~\cite{mcmahan2017communication}.
A common challenge in federated learning is client heterogeneity, as it complicates model convergence~\cite{xu2023asynchronous,ye2023heterogeneous,nishio2019client}.
Nonetheless, this issue remains a blind spot in current \gls{foltr} research.
Evaluations are typically based on IID random splits of traditional offline LTR datasets~\cite{wang2021efficient,kharitonov2019federated,wang2021effective,wang2021federated}.
However, this oversimplifies real-world settings where clients are heterogeneous~\cite{wang2022non,cecchetti2024learning}.



Public click datasets are rare.
Prior studies have shown that ``anonymized'' user IDs can easily be deanonymized~\cite{narayanan2006break,barbaro2006face}.
Since then, companies have become more cautious about releasing new datasets.
For example, when Microsoft released ORCAS~\cite{craswell2020orcas}, a click dataset derived from search interactions on Bing\footnote{\url{https://bing.com}}, they only included queries submitted by at least $k$ users, and removed user IDs and timestamps.
Researchers at Yandex~\cite{yandex-personalized-web-search-challenge} took a different approach by masking query terms and URLs, replacing them with numeric IDs.
Obfuscation techniques like this are effective at preserving user privacy but make it difficult to extract meaningful features for LTR.
The AOL dataset~\cite{pass2006picture}, to the best of our knowledge, remains the only publicly available source of raw query logs.
This work is not the first to attempt to reconstruct search result lists from AOL query logs.
Our method builds on the work of \citet{guo2021aol4ps}, who introduced AOL4PS. 
Their method was to leverage BM25 rankings over the document corpus in order to simulate result lists.
This approach was later picked up in simulations of OLTR in decentralized (peer-to-peer) settings~\cite{gregoriadis2025swarmsearch,gregoriadis2025decentralized}.
We extend this method with a random offset to debias the results, as well as the inclusion of ``natural candidates'', which we explain in \Cref{sec:method:natural}.
Furthermore, we use the Internet Archive\footnote{https://archive.org} to retrieve websites approximately at the time of the query logs (mid-2006).
Our corpus surpasses AOL4PS by \num{271392} documents, whilst also more faithfully reflecting the original state of the websites.

\begin{figure}[ht]
    \centering
    \includegraphics[width=\linewidth]{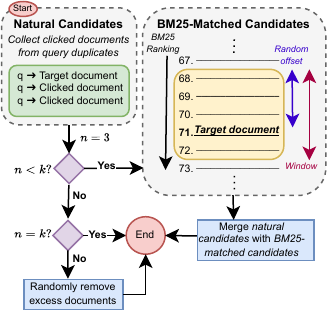}
    \caption{Example run of our top-k result list generation method. Natural candidates are extracted from activities with redundant queries, and then supplemented with BM25-matched candidates.}
    \label{fig:method}
\end{figure}

\section{Dataset Creation}\label{sec:method}

In 2006, AOL released a dataset of query logs from users of their web search engine~\cite{pass2006picture}.
To this date, it remains the only publicly available dataset that combines user identifiers, raw search queries, and the URL of the clicked document~\cite{macavaneysigir2021irds}.
More recently, \citet{macavaney2022reproducing} successfully restored the contents of the majority of documents using the Internet Archive.
Their approach ensured that the recovered content approximates the state of the documents at the time of the query.
Furthermore, they parsed the HTML, extracting title and body as plain text.
This forms the basis for our dataset.
Training and evaluation of a \gls{ltr} model requires a set of candidate documents for each query, i.e., the set of search results from which the user could choose.
In \gls{oltr}, this means that for every clicked document, we must also know the documents the user has decided \emph{against}.
This data is missing from the original dataset.
The reconstruction of result lists forms the core of our methodology.
After reconstructing result lists, we filter our dataset to only include the top \num{10000} users by number of query logs.

\subsection{Reconstructing Result Lists}

It is not possible to restore the result lists originally presented to users for each query.
To address this limitation, we simulate top-20 result lists for each query log based on the corpus of documents in the dataset. 
Our approach exploits redundant queries to extract \emph{natural candidates} and then supplements them with a BM25-based matching strategy inspired by \citet{guo2021aol4ps}.
We illustrate our approach in \Cref{fig:method}, and detail its components in the following.


\subsubsection{Natural candidates}\label{sec:method:natural}

Each \emph{query log} comprises the raw search query, and the clicked document (or \emph{target document}), among other metadata.
By the raw search query, it is often possible to identify duplicates across multiple query logs 
where the same search query was used and distinct documents were clicked.
For \qty{9.3}{\%} of queries, we could identify at least five distinct clicked documents; 
for \qty{44.3}{\%} of queries, it was at least two.
We define \emph{natural candidates} for a given query log as all documents clicked in any other query log with the same query, including the target document of the current query log. 
This definition is based on the assumption that if a document was clicked in one query log, it was also included in the candidate set for \emph{all} query logs corresponding to the same query.
We acknowledge that this is a strong assumption, as search engines often vary result lists based on factors such as user location, language preferences, personalization signals, and temporal dynamics.
However, we believe these issues are mitigated by the fact that the query logs were collected over a relatively short period of three months and exclusively from users located in the United States.
Moreover, personalized search in 2006 was only beginning to emerge around that time, and was far less advanced than it is today~\cite{dou2007large}.

\subsubsection{BM25-matched candidates}

We aimed for exactly $k=20$ candidates for each query log.
Usually, the number of natural candidates $n<20$.
In case of $n>20$, we randomly remove excess documents from the candidate set, but never the target document of the query log itself.
For the missing $k-n$ candidates, our approach leverages BM25 retrieval\footnote{BM25 is a standard ranking function based on keyword matching~\cite{robertson1995okapi,robertson2009probabilistic}.}.
We used \texttt{pyserini}~\cite{Lin_etal_SIGIR2021_Pyserini} to build an index of documents based on their title, body text, and URL.
Based on the BM25 search utility in this library, we generated a top-1000 ranking of documents matching our query.
We explicitly excluded natural candidates from this list, except for the target document.
A naive approach would be to select the top-ranking items to supplement the candidate list.
We observed that, most of the time, the target document is not within the top-k documents.
This could create an unintended bias that the ranker might learn during training.
To avoid such bias, therefore, we apply a window around the target document's position within the top-1000 ranking.
We set the window size $w=k-n+1$, to account for the missing candidates and the target document itself.
The windowing approach is inspired from \citet{guo2021aol4ps}, who also used it to reconstruct result lists in the AOL dataset.
Rather than centering the target document within the window, however, we placed it at a random offset between 0 and $w-1$. This strategy is again intended to mitigate potential sources of bias.

\subsection{Feature Selection}

After reconstructing result lists, we compiled query-document feature vectors according to the standard LETOR format~\cite{qin2013introducing}.
For each candidate document in each query log, we created a training record consisting of a query ID, a binary relevance label (1 if the candidate corresponds to the target document, 0 otherwise), and the feature vector.
The feature vector encodes all information used by the ranker to make relevance decisions. Consequently, identifying the most informative features is a critical aspect of LTR~\cite{geng2007feature}.
For AOL4FOLTR, we followed the conventions established by classic LTR datasets. Specifically, we replicated all features from WEB30k~\cite{qin2013introducing} that could be derived from our data.
This excludes features like PageRank or dwell time, over which we have no information.
In total, we employ 103 features.
Our feature selection is documented in our code repository.
Furthermore, our open-source approach and provision of raw data allow researchers to experiment by creating and adding new features.

\begin{figure}[t]
    \centering
    \input{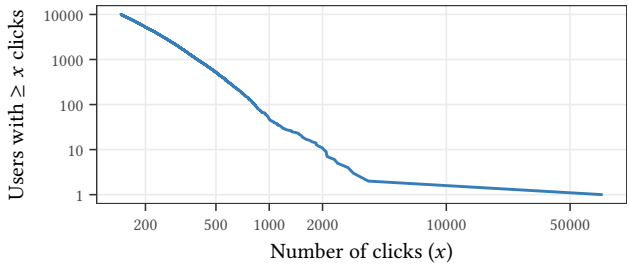}
    \vspace{-2em}
    \caption{Cumulative number of users by minimum click count (log-log). A minority of users account for most clicks.}
    \label{fig:userQueries}
\end{figure}

\begin{figure}[t]
    \centering
    \input{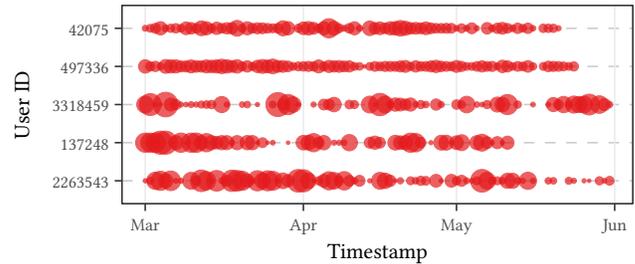}
    \vspace{-2em}
    \caption{Temporal activity patterns for a sample of users. User activities are bursty and irregular. Each dot represents activity on a given day, with dot size indicating the volume of activity.}
    \label{fig:queryBursts}
\end{figure}

\section{Dataset Analysis}\label{sec:analysis}

Our dataset comprises \num{2594705} query logs (\num{637996} unique queries) from \num{10000} unique users, who clicked on \num{428157} distinct documents.

\textbf{Data quantity.}
As is typical in such datasets, click activity exhibits power-law characteristics, with most clicks generated by a few users. We display this in \Cref{fig:userQueries}.

\textbf{Temporal patterns.}
User activity over time is highly variable and irregular.
Typically, users engage in short periods of concentrated interactions (often conceptualized as \emph{sessions}~\cite{macavaney2022reproducing}),
resulting in activity bursts.
In \Cref{fig:queryBursts}, we visualize these bursts for the five most active users in our dataset\footnote{User \texttt{71845} was excluded due to anomalously high activity.}.
Each dot represents user activity (i.e., clicks) on a given day, with the dot size indicating the number of clicks.

\begin{figure}[ht]
    \centering
    \input{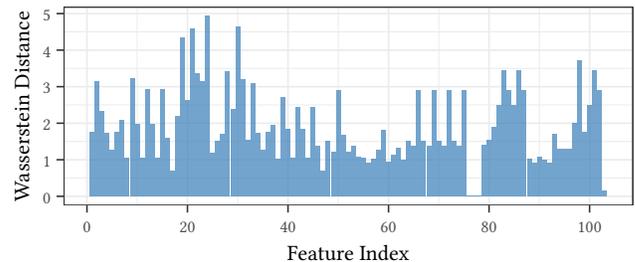}
    \vspace{-2em}
    \caption{Feature distribution divergence of clicked documents across clients.}
    \label{fig:featDivergence}
\end{figure}

\textbf{Feature heterogeneity.}
Local data heterogeneity is a known and well-studied problem in federated learning~\cite{zhao2018federated,zhu2021federated}.
Specifically, in the case of online learning to rank, it implies collaborative learning from clients with dissimilar click preferences with regards to the features of a search result~\cite{wang2022non,cecchetti2024learning}.
We measured this divergence by comparing the feature-wise probability distribution of clicked search results between all users.
Results are shown in \Cref{fig:featDivergence}.
Wasserstein distance (also known as Earth Mover's Distance) is a standard algorithm to measure feature skew in non-IID federated learning~\cite{zhao2018federated,solans2024non}.

\section{Experimental Evaluation}\label{sec:exp}

We evaluate our dataset for its application in both synchronous and asynchronous FOLTR.
To this end, we employ 100 clients corresponding to the top 100 users by number of query logs in the dataset.
As a benchmark, we construct an IID variant by randomly distributing the query logs of the 100 users across all clients.
Finally, we use a temporal split, holding out the latest \qty{20}{\%} for testing.

Our experiment uses FPDGD~\cite{wang2021effective}, the state-of-the-art algorithm in FOLTR.
Each client $c$ trains a local model $\theta^c$ on every personal search interaction, i.e., each click results in a local model update.
After a certain number of model updates $n_c$, the updated model $\theta^c_{t+1}$ is sent to the server, where model updates from all $\vert C\vert$ clients are received synchronously.
The server then performs a weighted averaging of all updates, as shown in \Cref{eq:fpdgd}.
The weight is determined by the number of queries each client has processed relative to other clients.
In our experiment, we use the original implementation and hyperparameters used by the authors of FPDGD~\cite{wang2021effective}, including added noise for differential privacy and a constant number of queries per update $n_c=4$.

\begin{equation}
    \theta_{t+1} = \sum_{c=1}^{|C|} \frac{n_c}{n} \theta^c_{t+1}, \quad \text{where} \quad n = \sum_{c=1}^{|C|} n_c
\label{eq:fpdgd}
\end{equation}

Evaluations of FOLTR in the literature have focused on the setting of synchronous federated learning.
In real systems, which are asynchronous, users send updates at different frequencies and at different times.
Specifically, in FOLTR applications,
client model updates are expected once a client has completed a batch of interactions~\cite{kharitonov2019federated,wang2021effective}.
Asynchronous FOLTR, therefore, must be able to handle stale updates, as outdated gradients may not align with the current global model.
To this end, we employ FedAsync~\cite{xie2019asynchronous},
a standard algorithm for dealing with staleness in asynchronous federated learning.
In FedAsync, received local updates are weighted according to staleness, as shown in \Cref{eq:fedasync}.
Staleness is measured by the number of rounds $r$ since the client has synchronized with the global model.

\begin{equation}\label{eq:fedasync}
    \theta_{t+1} = \frac{1}{1+r} \theta^c_{t+1} \cdot \left(1-\frac{1}{1+r}\right)\theta_t
\end{equation}

\begin{figure}[ht]
    \centering
    \input{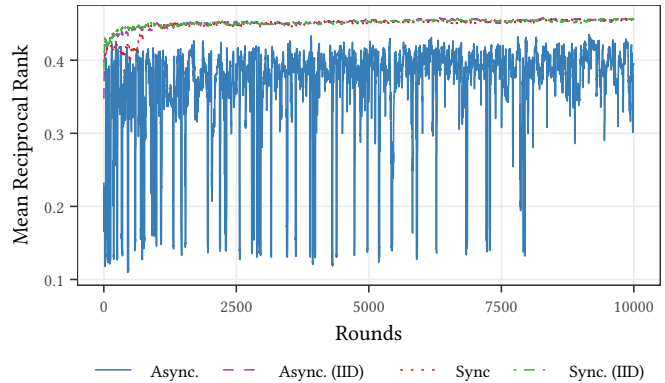}
    \vspace{-2em}
    \caption{Experimental evaluation of our dataset with 100 clients simulating users with the most queries in our dataset.}
    \label{fig:experiment}
\end{figure}

In \Cref{fig:experiment}, we present results for both synchronous and asynchronous learning of the global model, as measured by Mean Reciprocal Rank (MRR)\footnote{MRR is a standard metric when evaluating ranking quality~\cite{mcfee2010metric}.} and evaluated on the test set after each round.
We stopped the experiment after \num{10000} rounds.
In both settings, updates of individual clients are processed in chronological order, and each client update represents a batch of the client's $n_c$ ``next'' queries.
The \emph{synchronous} setting, however, does not respect the global order of client updates.
This is because each round $t$ processes the $t$th batch of all clients, despite timelines across clients not aligning.
When a client runs out of batches, it gets skipped.
In the \emph{asynchronous} setting, a round processes a single client update, and client updates are processed in the chronological order given by the timestamp of the last query in the batch.
A client's model only synchronizes with the global model after it sent its update.
That is, in the synchronized setting, the client models are updated after every round;
in the asynchronized setting, a client's model remains stale until the same client sends their next update.

Our experimental results indicate major performance instability in the asynchronous setting when simulating real user profiles.
This effect is entirely absent from our IID benchmark, which also exhibits higher overall MRR.
Further, we notice only minor differences with the IID benchmark in the synchronous setting, with convergence occurring after around \num{1000} rounds.
We hypothesize that incorporating features reflecting document content, beyond abstract keyword-matching metrics, may yield more pronounced differences, as suggested by the findings of \citet{wang2022non}.

\section{Conclusion}\label{sec:conclusion}

We introduced AOL4FOLTR, a novel online learning-to-rank dataset based on real user clicks, encompassing user IDs, query timestamps, and raw query and document contents.
This resource sets a new benchmark for the simulation of heterogeneous and asynchronous federated learning settings.
Our experiments demonstrated the importance of simulations with real data rather than IID data, as is found in current literature.
Nevertheless, we believe the true implications extend beyond the results presented here.
By releasing raw queries and document contents, we empower researchers to derive new LTR features.
Because of the availability of our data and methods, this resource offers broader relevance, with utility in the study of LTR feature selection, personalization techniques, and federated or decentralized information retrieval.


\begin{acks}
This work was funded by the Dutch National NWO/TKI Science Grant BLOCK.2019.004.
\end{acks}

\section*{GenAI Usage Disclosure}


Generative AI tools were used to assist with writing and coding for this project. All outputs were thoroughly reviewed by the authors, who take full responsibility for the content and integrity of this work.

\bibliographystyle{ACM-Reference-Format}
\bibliography{ref}


\end{document}